\documentclass[12pt]{article}

\usepackage{graphicx}
\usepackage{dsfont}
\usepackage{amssymb}
\usepackage{amsmath}
\usepackage{mathrsfs}

\begin{document}
\title{Perfect imaging without negative refraction}
\author{Ulf Leonhardt\\
School of Physics and Astronomy, University of St Andrews,\\
North Haugh, St Andrews KY16 9SS, UK
}
\date{\today}
\maketitle
\begin{abstract}
Perfect imaging has been believed to rely on negative refraction, but here we show that an ordinary positively--refracting optical medium may form perfect images as well. In particular, we establish a mathematical proof that Maxwell's fish eye in two--dimensional integrated optics makes a perfect instrument with a resolution not limited by the wavelength of light. We also show how to modify the fish eye such that perfect imaging devices can be made in practice. Our method of perfect focusing may also find applications outside of optics, in acoustics, fluid mechanics or quantum physics, wherever waves obey the two--dimensional Helmholtz equation. 
\\

\end{abstract}

\newpage

\section{Introduction}

The quest for the perfect lens \cite{Pendry} initiated and inspired the rise of research on metamaterials, because perfect imaging has been believed to rely on negative refraction \cite{Veselago}, an optical property not readily found in natural materials. Metamaterials may be engineered to exhibit negative refraction \cite{SPW,SLW}, but in such cases they tend to be absorptive and narrowband, for fundamental reasons \cite{Stockman}. Perfect optical instruments without the physical problems of negative refraction have been suggested long before \cite{Maxwell,Lenz,Stettler,Luneburg,BornWolf}, but they were only proven to be perfect for light rays, but not necessarily for waves. Here we establish a mathematical proof that the archetype of the perfect optical instrument, Maxwell's fish eye \cite{Maxwell}, is perfect by all standards: it has unlimited resolution in principle. We also show how to modify the fish eye for turning it into a perfect imaging device that can be made in practice, with the fabrication techniques that were applied for the implementation \cite{Valentine,Gabrielli,Lee} of optical conformal mapping \cite{Leo,LeoNotes,LiPendry,LeoNatMat}. Such devices may find applications in broadband far--field imaging with a resolution that is only limited by the substructure of the material, but no longer by the wave nature of light. 

Given the phenomenal interest in imaging with negative refraction, it seems astonishing how little attention was paid to investigating the previously known ideas for perfect optical instruments without negative refraction, in particular as they are described in {\it Principles of Optics} by Born and Wolf \cite{BornWolf}. The most famous perfect optical instrument, Maxwell's fish eye \cite{Maxwell}, was treated with Maxwell's equations \cite{Tai,Rosu} but without focusing on its imaging properties, and the same applies to the fish eye for scalar waves \cite{Makowski} and numerical simulations of wave propagation in truncated fish eyes \cite{Greenwood}. Here we analyse wave--optical imaging in a two--dimensional Maxwell fish eye, primarily because such a device can be made in integrated optics on a silicon chip for infrared light \cite{Valentine,Gabrielli} or possibly with gallium--nitride or diamond integrated optics for visible light. We begin our analysis with a visual exposition of the main ideas and arguments before we apply analytical mathematics to prove our results. 

\section{Visualisation}

Maxwell \cite{Maxwell} invented a refractive--index profile where all light rays are circles and, according to his paper, ``all the rays proceeding from any point in the medium will meet accurately in another point''. As Maxwell wrote, ``the possibility of the existence of a medium of this kind possessing remarkable optical properties, was suggested by the contemplation of the crystalline lens in fish'' --- hence fish eye --- ``and the method of searching for these properties was deduced by analogy from Newton's {\it Principia}, Lib.\ I. Prop.\ VII.''
Luneburg \cite{Luneburg} represented Maxwell's fish--eye profile in a beautiful geometrical form: the fish eye performs the stereographic projection of the surface of a sphere to the plane (or the 3D surface of a four--dimensional hypersphere to three-dimensional space).
As the surface of a sphere is a curved space --- with constant curvature --- the fish eye performs, for light, a transformation from a virtual curved space into physical space \cite{LeoPhilbin}; it is the simplest element of non--Euclidean transformations optics \cite{LeoTyc}, suggested for achieving broadband invisibility \cite{LeoTyc,Tycetal}.

\begin{figure}[h]
\begin{center}
\includegraphics[width=26.0pc]{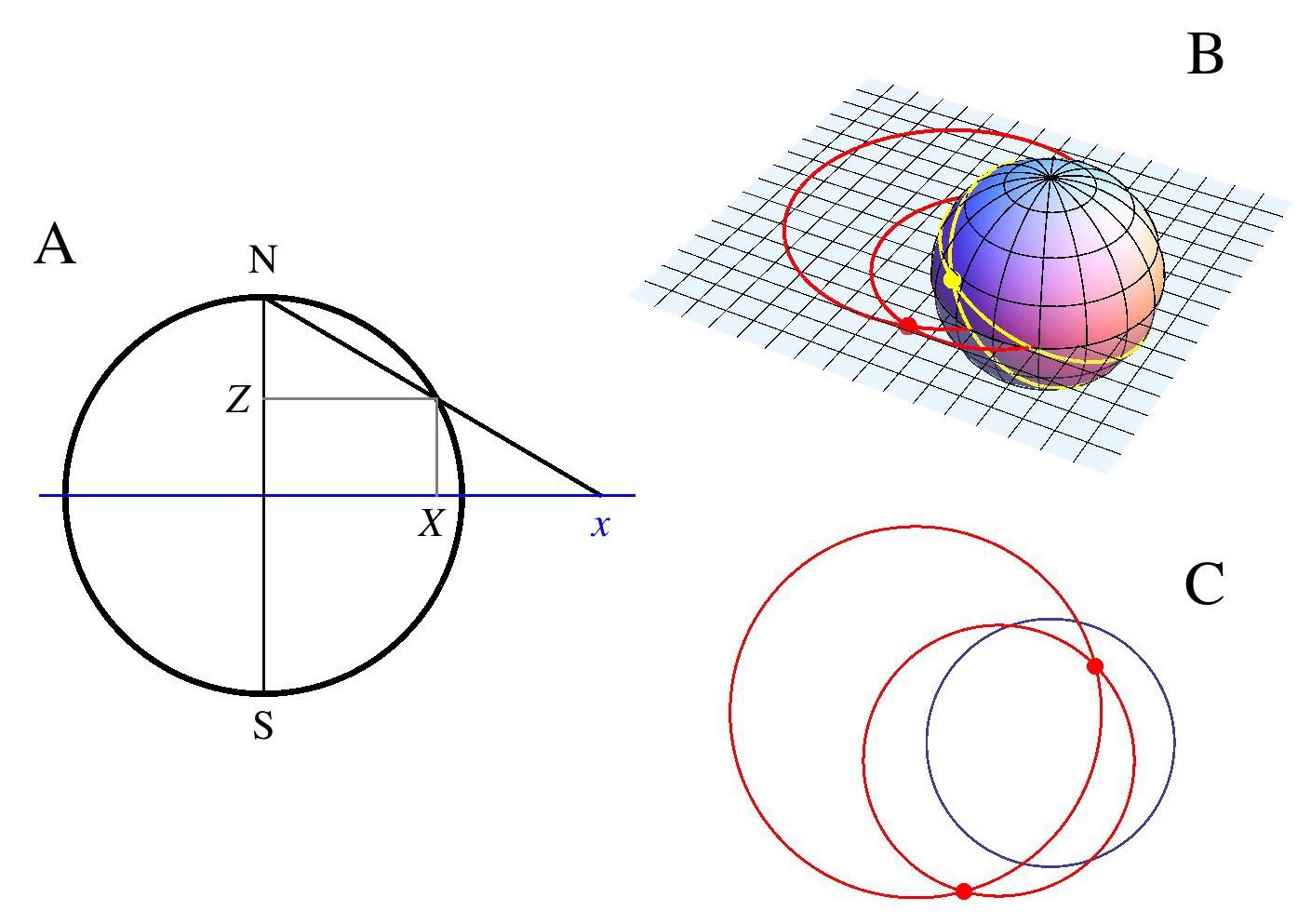}
\caption{
\small{
Stereographic projection. A: through each point $\{X,Y,Z\}$ on the surface of the sphere a line is drawn from the North Pole N. Where this line intersects the plane through the Equator lies the projected point  $\{x,y\}$. ($Y$ and $y$ not shown here.) The Northern Hemisphere is mapped to the exterior of the Equator with N at $\infty$, while the Southern Hemisphere is mapped to the interior; the South Pole S appears at the origin. B: circles on the sphere are projected into circles on the plane. Light rays, shown in yellow on the sphere and in red on the plane, are the geodesics, the great circles on the sphere. All the rays emitted from one point meet again at the antipodal point, forming a perfect image there. C illustrates how the light circles meet in the stereographic projection with the Equator shown in blue.
}
\label{fig:stereo}}
\end{center}
\end{figure}

The stereographic projection, invented by Ptolemy, lies at the heart of the Mercator projection \cite{Needham} used in cartography. Figure \ref{fig:stereo} shows how the points on the surface of the sphere are projected to the plane cut through the Equator. Through each point, a line is drawn from the North Pole that intersects this plane at one point, the projected point. In this way the surface of the sphere is mapped to the plane and vice versa; both are equivalent representations. In the following we freely and frequently switch between the two pictures, the sphere and the plane, to simplify arguments.

Imagine light rays on the surface of the sphere. They propagate along the geodesics, the great circles. Consider a bundle of light rays emitted at one point. All the great circles departing at this point must meet again at the antipodal point on the sphere, see Fig.\ \ref{fig:stereo}B. The stereographic projection maps circles on the sphere to circles on the plane \cite{Needham}. Consequently, in an optical implementation of the stereographic projection \cite{Luneburg} --- in Maxwell's fish eye \cite{Maxwell} --- all light rays are circles and all rays from one point meet at the projection of the antipodal point, creating a perfect image.

\begin{figure}[h]
\begin{center}
\includegraphics[width=23.0pc]{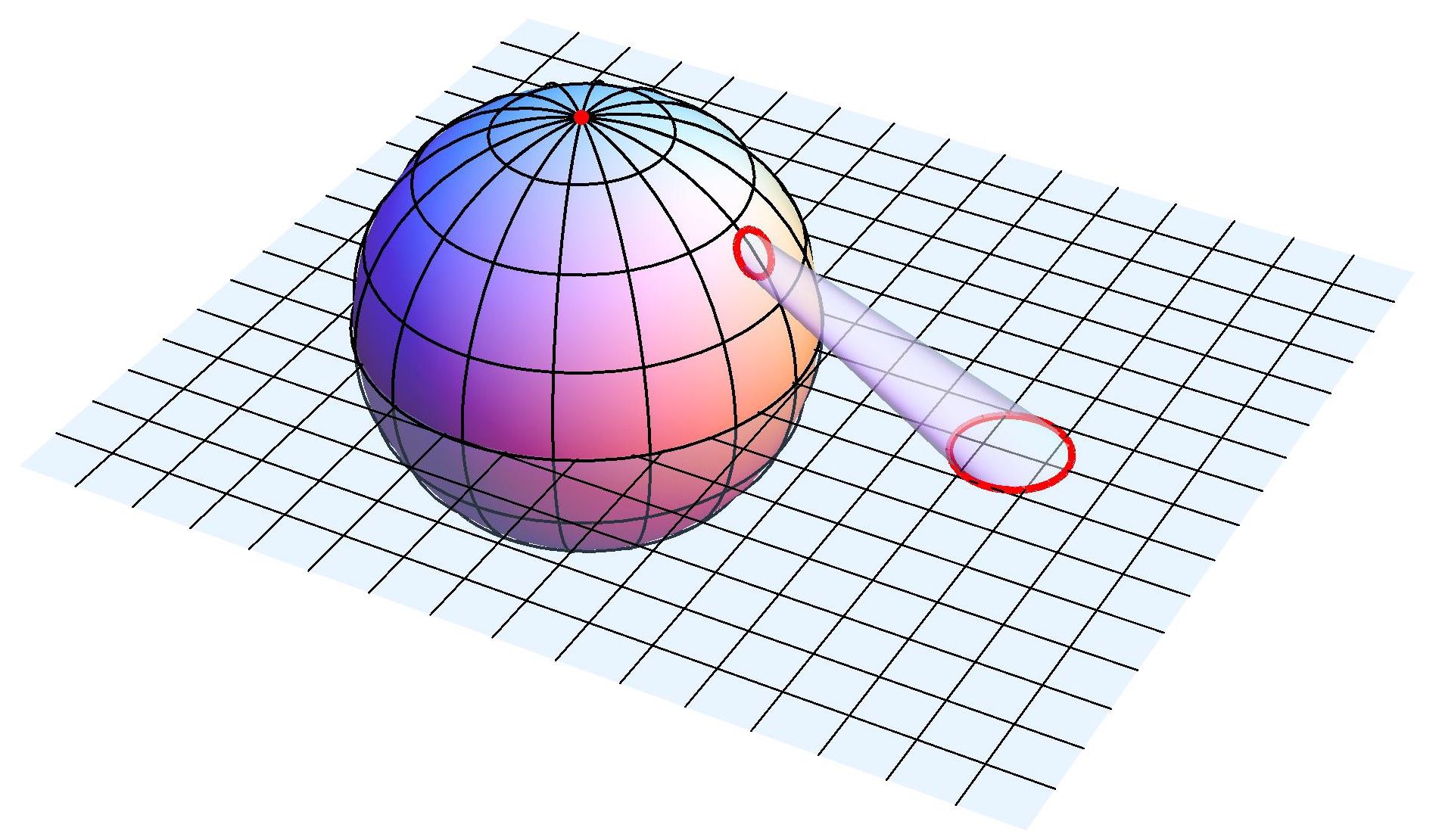}
\caption{
\small{
Isotropic medium. To implement the stereographic projection, the refractive index of the optical medium must be given by the ratio between a line element on the sphere and the corresponding projected line element. As the stereographic projection maps circles into circles, this ratio cannot depend on the direction of the line element: the medium is isotropic (the stereographic projection is a conformal map). The figure shows how circles on the Northern Hemisphere are magnified, requiring refractive indices below the index on the sphere, $n_1$. Circles on the Southern Hemisphere (not shown) are reduced by maximally a factor of two; the index ranges from $n_1$ to $2n_1$ in the interior of the Equator. Maxwell's fish eye with the index profile (\ref{maxwell}) turns out to perform the stereographic projection, see Eq.\ (\ref{elements}).
}
\label{fig:circles}}
\end{center}
\end{figure}

If one wishes to implement the stereographic projection, creating the illusion that light propagates on the surface of a sphere, one needs to make a refractive--index profile $n$ in physical space that matches the geometry of the spherical surface with uniform index $n_1$. The refractive index $n$ is the ratio between a line element in virtual space and the corresponding line element in physical space \cite{LeoTyc}. In general, this ratio depends on the direction of the line element and so the geometry--implementing material is optically anisotropic. However, as the stereographic projection transforms circles into circles, even infinitesimal ones, the ratio of the line elements cannot depend on direction: the medium is optically isotropic, see Fig.\ \ref{fig:circles}. From this figure we can read off the essential behavior of the required index profile $n$. At the Equator, $n$ is equal to the index $n_1$ of the sphere. At the projection of the South Pole, the origin of the plane, $n$ must be $2 n_1$. We also see that $n$ tends to zero near the projection of the North Pole, infinity. Maxwell's exact expression for the fish--eye profile \cite{Maxwell} that performs the stereographic projection \cite{Luneburg} interpolates through these values, 
\begin{equation}
n = \frac{2 n_1}{1+r^2} \,.
\label{maxwell}
\end{equation}
Here $r$ denotes the distance from the origin of the plane measured in terms of the size of the device. In these dimensions, the Equator lies at the unit circle. Beyond the Equator, in the projected region of the Northern Hemisphere with $r>1$, the index falls below $n_1$ and eventually below 1; the speed of light in the material must exceed the speed of light in vacuum.

\begin{figure}[h]
\begin{center}
\includegraphics[width=26.0pc]{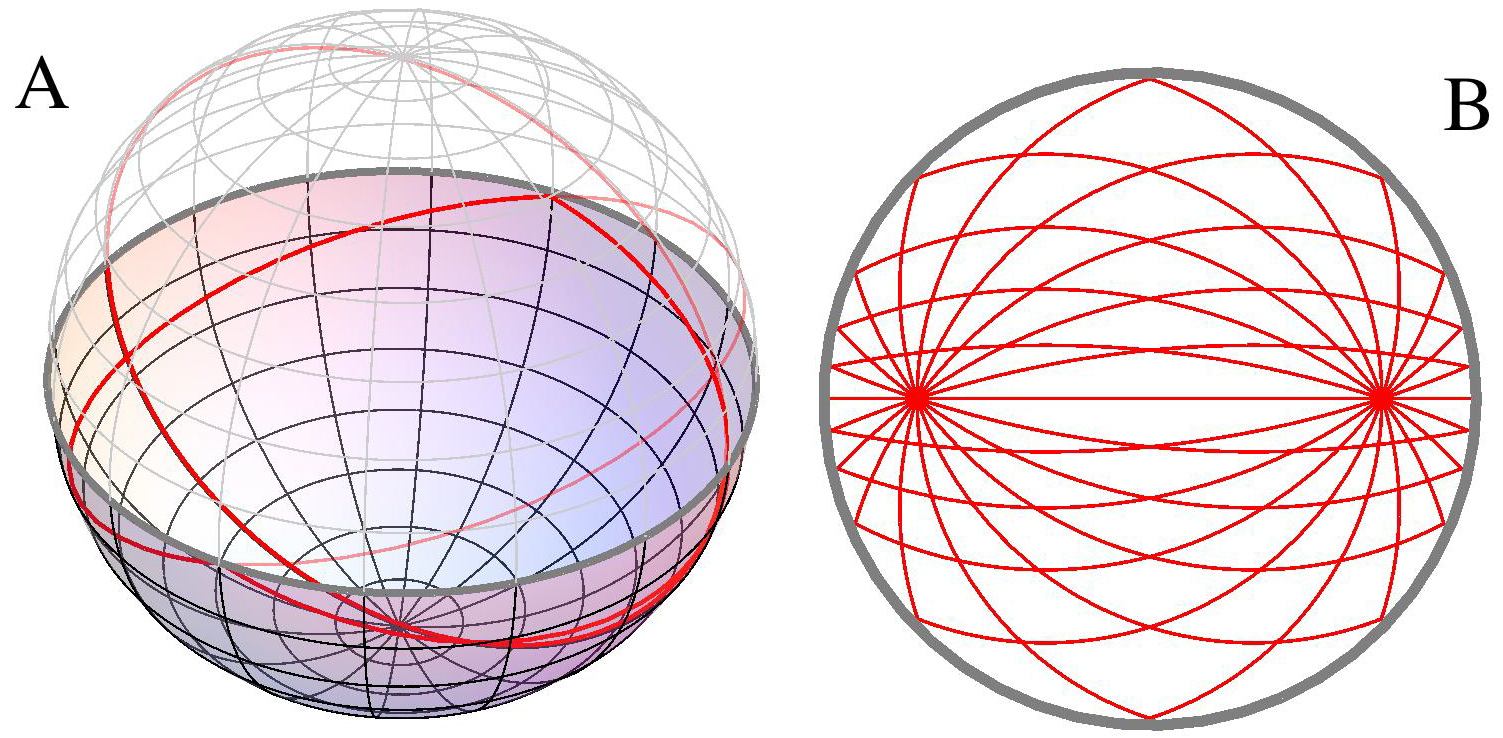}
\caption{
\small{
Fish--eye mirror. A: a mirror at the Equator of the sphere creates the illusion that light rays, shown in red, perform complete great circles, whereas in reality they are reflected. Picture B shows light rays emitted from one point on the plane in the stereographic projection performed by Maxwell's fish eye. The reflected rays from an arbitrary point all meet at the corresponding image point.
}
\label{fig:mirror}}
\end{center}
\end{figure}

In order to avoid the apparent need for superluminal propagation, we adopt an idea from non--Euclidean cloaking \cite{LeoTyc}:
imagine we place a mirror around the Equator, see Fig.\ \ref{fig:mirror}A. For light propagating on the Southern Hemisphere, the mirror creates the illusion that the rays are doing their turns on the Northern Hemisphere, whereas in reality they are reflected. Figure \ref{fig:mirror}B shows that the reflected image of the antipodal point is the mirror image of the source in the plane (an inversion at the centre). Each point within the mirror--enclosed circle creates a perfect image. In contrast, an elliptical mirror has only two focal points, instead of focal regions, and is therefore less suitable for imaging.

The required refractive--index profile (\ref{maxwell}) for $r \leq 1$ can be manufactured on planar chips, for example by diluting silicon with air holes \cite{Valentine} or by enhancing the index of silica with pillars of silicon \cite{Gabrielli}. The index contrast $n(0) / n(1)$ of $2$ is achievable for infrared light around $1500$nm. Gallium--nitride or diamond integrated optics could be used to create suitable structures for visible light. Such devices may be employed for transferring images from a nano--stamp or in other applications, provided the image resolution is significantly better than the wavelength. In the following we show that this is indeed the case. 

\begin{figure}[h]
\begin{center}
\includegraphics[width=26.0pc]{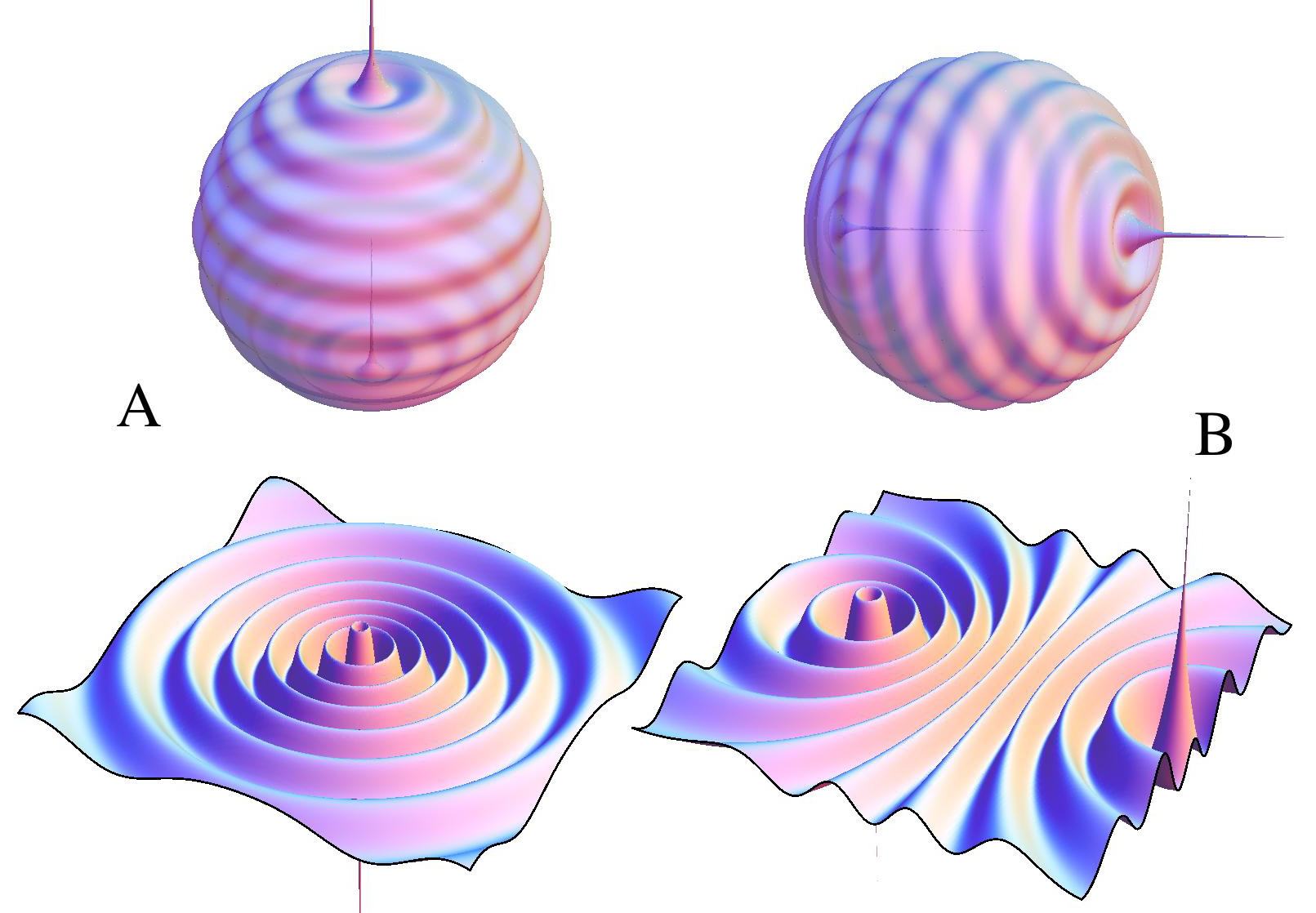}
\caption{
\small{
Light waves. A: the wave emitted by a source at the South Pole is visualized on the sphere and projected to the plane. The pictures show the real part of the electric field (\ref{prep}) with $\nu=20.25$; for the visualization on the sphere the radius is modulated as $1+0.5\,\mathrm{Re}E_\nu$. At the North Pole the field focuses to a point. B: in order to establish the wave emitted at an arbitrary source point, the field shown in A is rotated on the sphere and then projected. On the plane, the rotation corresponds to the rotational M\"obius transformation (\ref{rot}) with angles $\gamma$ and $\chi$. The rotation angle of the sphere is $2\gamma$ (with $\gamma=-0.2\pi$ here). The pictures illustrate the needle--sharp imaging in Maxwell's fish eye.
}
\label{fig:green}}
\end{center}
\end{figure}

It is sufficient to establish the electromagnetic field of a point source with unit strength, the Green's function, because any other source can be considered as a continuous collection of point sources with varying densities; the generated field is a superposition of the Green's functions at the various points. 
First, we deduce the Green's function for the most convenient source point, the origin, the stereographic projection of the South Pole. We expect from the symmetry of the sphere that the electromagnetic wave focuses at the North Pole as if it were the source at the South Pole in reverse, and this is what we prove in the next section. The field at the South Pole thus is a perfect image of the source field at the North Pole. Figure \ref{fig:green}A illustrates this Green's function.
Then we take advantage of the rotational symmetry of the sphere and rotate the point source with its associated electromagnetic field on the sphere, see Fig.\ \ref{fig:green}B. The stereographic projection to the plane gives the desired Green's function for an arbitrary source point. As the field is simply rotated on the sphere, we expect perfect imaging regardless of the source, which we prove in the next section as well.

\begin{figure}[h]
\begin{center}
\includegraphics[width=26.0pc]{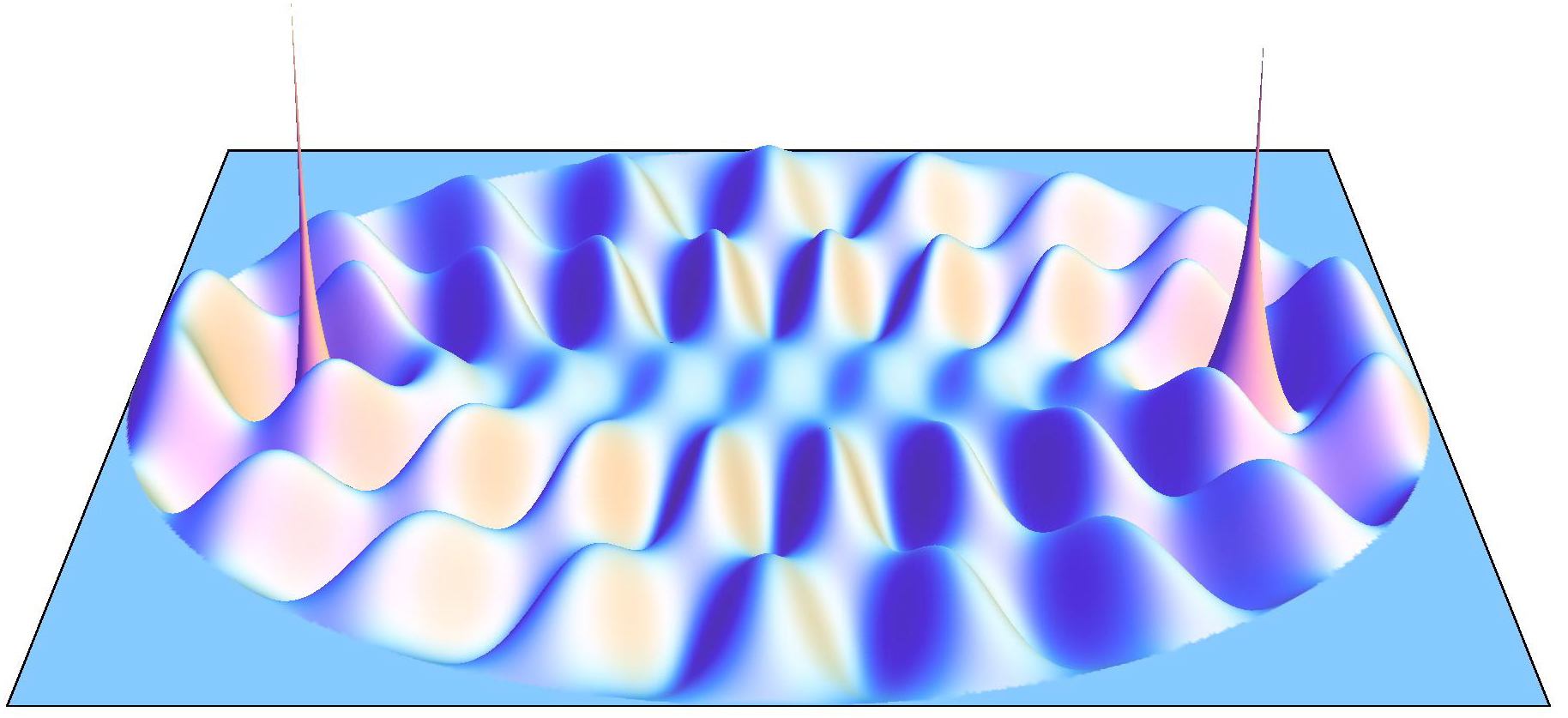}
\caption{
\small{
Imaging in the fish--eye mirror. The infinitely sharp field emitted at the source point (left tip) propagates as an electromagnetic wave until it focuses at the image point (right tip) with infinite resolution. The focusing is done by Maxwell's fish eye constrained by the mirror of Fig.\ \ref{fig:mirror}. For better visibility, the figure shows $-\mathrm{Re}E_k$, given by Eq.\ (\ref{field}) with the parameters of Fig.\ \ref{fig:green}. The image carries the phase shift $\nu\pi$ and has the opposite sign of the image in the fish eye without mirror.
}
\label{fig:imaging}}
\end{center}
\end{figure}

Finally, we include the reflection at the mirror, essentially by applying an adaptation of the method of images in electrostatics \cite{Jackson}. Figure \ref{fig:imaging} shows the result: Maxwell's fish eye, constrained by a mirror, makes a perfect lens.

\section{Calculation}

In this section we substantiate our visual arguments by analytical mathematics. It is convenient to use complex numbers $z = x + \mathrm{i} y$ for representing the Cartesian coordinates $x$ and $y$ of the plane. In the stereographic projection \cite{Needham}, the points $\{X,Y,Z\}$ on the surface of the unit sphere are mapped into
\begin{equation}
z = \frac{ X + \mathrm{i} Y}{1-Z}      \,,
\label{stereo}
\end{equation}
or, in spherical coordinates $\theta$ and $\phi$ on the sphere,
\begin{equation}
z = \mathrm{e}^{\mathrm{i} \phi} \cot \left( \frac{\theta}{2} \right)   \,.
\label{stereosphere}
\end{equation}
We obtain for Maxwell's index profile (\ref{maxwell}) with $r=|z|$ the formula
\begin{equation}
n = 2 n_1 \sin^2 \left( \frac{\theta}{2} \right)   \,.
\end{equation}
We express the line elements $\mathrm{d}x$ and $\mathrm{d}y$ in terms of the spherical coordinates and find 
\begin{equation}
n^2 \left( \mathrm{d}x^2 + \mathrm{d}y^2 \right) =   n_1^2 \left( \mathrm{d}\theta^2 + \sin^2 \theta \, \mathrm{d}\phi^2 \right)   \,.
\label{elements}
\end{equation}
The line element on the sphere thus differs from the Cartesian line element in the plane by the ratio $n/n_1$, a conformal factor that modifies the measure of length, but not the measure of angle: the stereographic projection is a conformal map \cite{Needham}. The optical medium (\ref{maxwell}) that implements this map is isotropic. Equation (\ref{elements}) proves \cite{Luneburg} that Maxwell's fish eye (\ref{maxwell}) indeed performs the stereographic projection (\ref{stereo}).

Consider TE--polarized light \cite{LL8} where the electric--field vector $\mathbf{E}$ points orthogonal to the plane.
In this case, only one vector--component $E$ matters, the orthogonal component. By Fourier analysis, we expand $E$ in terms of monochromatic fields $E_k$. They obey the Helmholtz equation \cite{BornWolf,LL8}
\begin{equation}
\left( \nabla^2 + n^2 k^2 \right) E_k =0     \, ,
\label{helmholtz}
\end{equation}
except at the source and image points.
Close to the source point, $E_k$ should approach the logarithmic field of a line source \cite{Jackson}. We also require that the field $E$ is retarded, {\it i.e.}\ in the time domain
\begin{equation}
E (z,t) = \int_{-\infty}^{+\infty} E_k \, \mathrm{e}^{- \mathrm{i} k t} \, \mathrm{d}k = 0  \quad \mathrm{for} \quad t<0  \,,
\label{fourier}
\end{equation}
where $t$ denotes time in appropriate units (of propagation length measured in terms of the dimensions of the device).
For simplicity, we rescale the wave number $k$ such that
\begin{equation}
n_1 = 1  \,.
\end{equation}
If we manage to show that the field $E_k$ is also logarithmic near the image point, as if the image were a line source run in reverse, an infinitely--well localized drain, we have proven perfect imaging with unlimited resolution. We need to supplement the optical medium with a drain as well as a source, for the following reason. In writing down the Helmholtz equation (\ref{helmholtz}) we consider monochromatic waves in a stationary state. However, the source is continuously generating a stream of electromagnetic waves that must disappear somewhere. In free space, the waves would disappear in the infinitely far distance, at infinity. In the case of imaging, the waves must find a finite drain, for otherwise a stationary state cannot exist. We must assume that the waves are absorbed at the image. However, source and drain ought to be causally connected as well; in our model we cannot simply place an arbitrary inverse source at the expected image point. The field at the drain must exhibit a phase shift due to the time delay between source and image, and the Green's function must be retarded according to Eq.\ (\ref{fourier}). Causality and infinite resolution are both required for proving perfect imaging. 

Consider the case of the most convenient source point, illustrated in Fig.\ \ref{fig:green}A. Here the source is placed at the origin, the stereographic projection of the South Pole. As it is natural to assume radial symmetry, the Helmholtz equation (\ref{helmholtz}) is reduced to
\begin{equation}
\frac{1}{r} \frac{\partial}{\partial r} \, r \, \frac{\partial E_k}{\partial r} +
n^2 k^2 E_k = 0
\end{equation}
in polar coordinates with $r=|z|$.
The general solution of this ordinary differential equation is a superposition of Legendre functions \cite{Erdelyi} $P_{\nu} (\pm \zeta)$ with the index
\begin{equation}
\nu = \frac{1}{2} \left( \sqrt{4 k^2 + 1} - 1 \right) \quad \mathrm{or, equivalently,} \quad  k^2 = \nu (\nu+1) 
\label{nu} 
\end{equation}
and the variable
\begin{equation}
\zeta = \cos \theta = \frac{r^2 - 1}{r^2 + 1}   \,.
\end{equation}
The relation between $k$ and $\nu$ is the same as between the wavenumber and index of a spherical harmonic, 
but $\nu$ is not necessarily an integer.
Let us write down the specific solution
\begin{equation}
E_{\nu} = \frac{P_{\nu} (\zeta) -  \mathrm{e}^{\mathrm{i} \nu \pi} \, P_{\nu} (-\zeta) }{4 \sin(\nu\pi)} 
\label{prep} 
\end{equation}
that can also be expressed in terms of the Legendre function $Q_{\nu}$ of the second kind \cite{Erdelyi}, 
\begin{equation}
E_{\nu} = \frac{\mathrm{e}^{\mathrm{i} \nu \pi}}{2 \pi } \, Q_{\nu} (\zeta)   \,. 
\label{qrep} 
\end{equation}
Note that the definition of $Q_{\nu}$ is sometimes ambiguous --- it depends on the branch chosen on the complex plane --- and so we generally prefer the expression (\ref{prep}) here.
Note also that this expression has a meaningful limit for integer $\nu$ when both the denominator and the numerator tend to zero \cite{Erdelyi}.
We obtain from Eqs.\ 3.9.(9) and 3.9.(15) of Ref.\ \cite{Erdelyi} the asymptotics
\begin{equation}
E_{\nu} \sim \frac{\ln r}{2 \pi }  \quad \mathrm{for} \quad r \rightarrow 0  \,,
\label{source} 
\end{equation}
which proves that formula (\ref{prep}) describes the electromagnetic wave emitted from a line source, because for $r$ within a small circle around the origin we get
\begin{equation}
\int \!\! \int \left( \nabla^2 + n^2 k^2 \right) \frac{\ln r}{2 \pi } \, \mathrm{d}A \sim
\int \!\! \int \frac{\nabla^2 \ln r}{2 \pi } \, \mathrm{d}A  =
\oint \frac{\nabla \ln r}{2\pi} \cdot \mathrm{d}\mathbf{s} = 1 \,,
\end{equation}
where we used Gauss' theorem with $\mathrm{d}\mathbf{s}$ pointing orthogonal to the integration contour.  
In order to prove that the Green's function (\ref{prep}) is retarded, we utilize the integral representation 3.7.(12) of Ref.\ \cite{Erdelyi} for $Q_{\nu}$ in expression 
(\ref{qrep}),
\begin{equation}
E_{\nu} = \frac{\mathrm{e}^{\mathrm{i} \nu \pi}}{2 \pi } \, \int_{0}^{+\infty} \frac{\mathrm{d}\xi}{\left( \cos \theta + \mathrm{i} \cosh \xi \sin \theta  \right)^{\nu+1}}    \,.
\label{int} 
\end{equation}
The Green's function thus is an integral over powers in $\nu$. As $\mathrm{arg} (\cos \theta + \mathrm{i} \cosh \xi \sin \theta)$ $\leq \pi$ for $0\leq\theta\leq\pi$ and $\nu \rightarrow k$ for $k \rightarrow \infty$, the integrand multiplied by $\exp(\mathrm{i} \nu \pi)$ falls off exponentially on the upper half of the complex $k$ plane. Therefore we can close the integration contour of the Fourier transform (\ref{fourier}) there. As the integrand of the representation (\ref{int}) multiplied by $\exp(\mathrm{i} \nu \pi)$ is analytic in $k$, the Fourier integral (\ref{fourier}) vanishes for $t<0$, which proves that $E_{\nu}$ describes the retarded Green's function. We also obtain from Eqs.\ 3.9.(9) and 3.9.(15) of Ref.\ \cite{Erdelyi} the asymptotics
\begin{equation}
E_{\nu} \sim \mathrm{e}^{\mathrm{i} \nu \pi} \, \frac{\ln r}{2 \pi }  \quad \mathrm{for} \quad r \rightarrow \infty  \,,
\label{image} 
\end{equation}
which proves that the image at infinity is perfectly formed, with a phase delay of $\nu\pi$. Furthermore, we get from Eq.\ 3.9.(2) of Ref.\ \cite{Erdelyi} the convenient asymptotic formula 
\begin{equation}
E_{\nu} \sim \mathrm{e}^{\mathrm{i} (\nu - 1/2) \pi} \, \frac{\Gamma (\nu+1)}{4 \,\Gamma (\nu+3/2)} \, \frac{(r-\mathrm{i})^{\nu+1} \, (r+\mathrm{i})^{-\nu}}{\sqrt{\pi r}} 
\label{asymp} 
\end{equation}
for large $\nu$ and $r$ located somewhere between $0$ and $\infty$, an excellent approximation for the Green's function (\ref{prep}), except near source and image.

So far we established the Green's function of a source at the origin. In order to deduce the Green's function of an arbitrary source point, we utilize the symmetry of the sphere in the stereographic projection illustrated in Fig.\ \ref{fig:green}. We rotate the source with its associated field from the South Pole (Fig.\ \ref{fig:green}A) to another, arbitrary point on the sphere and project it to the plane (Fig.\ \ref{fig:green}B). Rotations on the sphere correspond to a subset of M\"obius transformations on the complex plane \cite{Needham}. A M\"obius transformation is given by a bilinear complex function with constant complex coefficients,
\begin{equation}
z' = \frac{a \, z + b}{c \, z + d}
 \quad \mathrm{with} \quad a \, d - b \, c = 1  \,. 
\end{equation}
A rotation on the sphere corresponds to \cite{Needham} 
\begin{equation}
z' = \frac{z \cos \gamma - \mathrm{e}^{\mathrm{i} \chi} \sin \gamma}
{z \, \mathrm{e}^{-\mathrm{i} \chi} \sin \gamma + \cos \gamma }    \,. 
\label{rot} 
\end{equation}
We obtain for the Laplacian in the Helmholtz equation (\ref{helmholtz}) 
\begin{equation}
\nabla^2 = 4 \, \frac{\partial^2}{\partial z \partial z^*} = \left| \frac{\mathrm{d} z'}{\mathrm{d} z}  \right|^2 \, \nabla'^2 = | c \, z + d  |^{-4} \, \nabla'^2 \,.
\end{equation}
From the relations 
\begin{equation}
|a|^2 + |b|^2 = |c|^2 + |d|^2 = 1 \,,\quad
a \, b^* + c \, d^* = 0   
\end{equation}
for rotations (\ref{rot}) we get the transformation of the refractive--index profile (\ref{maxwell}) in the Helmholtz equation (\ref{helmholtz})
\begin{equation} 
| c \, z + d  |^2 \, n =  \frac{2 n_1 | c \, z + d  |^2}{1 + |z|^2} = \frac{2 n_1 | c \, z + d  |^2}{| a \, z + b |^2 + | c \, z + d |^2} =
\frac{2 n_1}{1 + |z'|^2}  \,.
\end{equation}
Consequently, for Maxwell's fish eye, the Helmholtz equation (\ref{helmholtz}) is invariant under rotational M\"obius transformations,
which simply reflects the rotational symmetry of the sphere in the stereographic projection. We see from the inverse M\"obius transformation
\begin{equation}
z = \frac{a \, z' - b}{- c \, z' + d}
\end{equation}
that the source point at $z'=0$ has moved to 
\begin{equation}
z_0 = - \frac{b}{d} = \mathrm{e}^{\mathrm{i} \chi} \tan \gamma
\end{equation}
and that the image at $z'=\infty$ appears at 
\begin{equation}
z_{\infty} = - \frac{a}{c} = - \, \mathrm{e}^{\mathrm{i} \chi} \cot \gamma = - \frac{1}{z_0^*}    \,.
\end{equation}
The electric field is given by the expression (\ref{prep}) with
\begin{equation}
r = |z'|   \,.
\end{equation}
Near the source $z_0$, where $z' \rightarrow 0$, we linearize the M\"obius transformation (\ref{rot}) in $z-z_0$ and near the image $z_{\infty}$, where $z' \rightarrow \infty$, we linearize $1/z'$ in $z-z_{\infty}$.
In the logarithmic expressions (\ref{source}) and (\ref{image}) the linearization prefactors just produce additional 
constants that do not alter the asymptotics. Consequently, 
\begin{equation}
E_{\nu} \sim \frac{\ln |z-z_0|}{2 \pi }  \, , \quad
E_{\nu} \sim - \, \mathrm{e}^{\mathrm{i} \nu \pi} \, \frac{\ln |z-z_{\infty}|}{2 \pi }    \,.
\label{si} 
\end{equation}
Maxwell's fish eye creates perfect images, regardless of the source point. The minus sign in the image field indicates that the electromagnetic wave emitted at $z_0$ with unit strength focuses at $z_{\infty}$ as if the image were a source of precisely the opposite strength. In addition, the image carries the phase delay $\nu \pi$ caused by the propagation in the index profile (\ref{maxwell}) or, equivalently, on the virtual sphere. Due to the intrinsic curvature of the sphere, the delay constant (\ref{nu}) is not linear in the wave number $k$, but slightly anharmonic. As the phase delay is uniform, however, a general source distribution is not only faithfully but also coherently imaged.

Finally, we turn to the wave optics of Maxwell's fish eye confined by a circular mirror, the case illustrated in Figs.\ \ref{fig:mirror} and \ref{fig:imaging}.
At the mirror, the electric field must vanish. Suppose we account for the effect of the mirror by the field of a virtual source, similar to the method of images in electrostatics \cite{Jackson}. The virtual source should have the opposite sign of the real source and, on the sphere, we expect it at the mirror image of the source above the plane through the Equator, at $\theta+\pi$ and $\phi+\pi$. The stereographic projection (\ref{stereosphere}) of the mirrored source is the inversion in the unit circle \cite{Needham}
\begin{equation}
z' = \frac{1}{z^*}  \,. 
\label{mirrortrans} 
\end{equation}
Consider the transformation (\ref{mirrortrans}), not only for the source, but for the entire electric field. We obtain for the Laplacian in the Helmholtz equation (\ref{helmholtz})
\begin{equation}
\nabla^2 = 4 \, \frac{\partial^2}{\partial z \partial z^*} = \left| \frac{\mathrm{d} z'}{\mathrm{d} z^*}  \right|^2 \, \nabla'^2 = |z|^{-4} \, \nabla'^2
\end{equation}
and for the transformed refractive--index profile (\ref{maxwell})
\begin{equation} 
|z|^2 \, n = \frac{2 n_1}{1 + |z'|^2}  \,.
\end{equation}
So the mirror image $E_{\nu} (1/z^*)$ of the field $E_{\nu} (z)$  is also a valid solution. Let us add to the field $E_\nu(z)$ of the original source the field $-E_\nu(1/z^*)$ of the virtual source conjured up by the mirror,
\begin{equation} 
E_k =  E_{\nu} (z) - E_{\nu} (1/z^*) \,.
\label{field}
\end{equation}
At the unit circle $1/z^*$ is equal to $z$, and so the field $E_k$ vanishes here. Consequently, formula (\ref{field}) satisfies the boundary condition and thus describes the correct Green's function of the fish--eye mirror. The image inside the mirror is the image of the virtual source. From the transformation (\ref{mirrortrans}) follows that the image point $z'_{\infty}$ is located at
\begin{equation} 
z'_{\infty} = - z_0 \,.
\end{equation}
We obtain from the formula (\ref{field}) and the asymptotics (\ref{si}) that
\begin{equation}
E_k \sim \mathrm{e}^{\mathrm{i} \nu \pi} \, \frac{\ln |z-z'_{\infty}|}{2 \pi }    \,.
\label{mirrorimage}
\end{equation}
The sign flip compared to Eq.\ (\ref{si}) results from the $\pi$ phase shift at the mirror, but the overall phase delay remains uniform, $(\nu+1)\pi$. The resolution is unlimited, and so the fish--eye mirror forms perfect images by all standards. The device may even tolerate some degree of absorption in the material. For example, assume that absorption appears as an imaginary part of the refractive index that is proportional to the dielectric profile. This case is equivalent to having an imaginary part of the wave number $k$ for the real refractive--index profile (\ref{maxwell}). Here we have established the Green's function for all $k$, including complex ones. As the asymptotics (\ref{mirrorimage}) is independent of $k$, apart from the prefactor that accounts for the loss in amplitude, such absorption does not affect the image quality. 

\section{Discussion}

Maxwell's fish eye \cite{Maxwell} makes a perfect lens; but it is a peculiar lens that contains both the source and the image inside the optical medium. Negatively--refracting perfect lenses \cite{Pendry} are ``short--sighted'' optical instruments, too, where the imaging range is just twice the thickness of the lens \cite{GREE}, but there source and image are outside the device. Hyperlenses \cite{Jacob,Liu} funnel light from microscopic objects out into the far field, for far--field imaging beyond the diffraction limit, but the resolution of a hyperlens is limited by its geometrical dimensions; it is not infinite, even in principle.

Fish--eye mirrors could be applied to transfer embedded images with details significantly finer than the wavelength of light over distances much larger than the wavelength, a useful feature for nanolithography. To name another example of potential applications, fish--eye mirrors could establish extremely well--defined quantum links between distant atoms or molecules embedded in the dielectric, for example colour centres in diamond \cite{DiamondCC}. Fish--eye mirrors could also find applications outside of optics, wherever waves obey the two--dimensional Helmhotz equation (\ref{helmholtz}) with a controllable wave velocity. For example, they could make ideal whispering galleries for sound waves or focus surface waves on liquids, or possibly create strongly entangled quantum waves in quantum corrals \cite{Heller}.

Like the negatively--refracting perfect lens \cite{Pendry} with electric permittivity and magnetic permeability set to $-1$, the fish--eye mirror does not magnify images. Note, however, that by placing the mirror at the stereographic projections of other great circles than the Equator, one could make magnifying perfect imaging devices. One can also implement, by optical conformal mapping \cite{Leo}, conformal transformations of fish eyes \cite{Luneburg} that form multiple images. As the fish--eye mirror consists of an isotropic dielectric with a finite index contrast, it can be made with low--loss materials and operate in a broad band of the spectrum. The image resolution is unlimited in principle. In practice, the dimensions of the sub--wavelength structures of the material will limit the resolution. If the required index profile (\ref{maxwell}) is created by doping a host dielectric one could perhaps reach molecular resolution.

In this paper, we focused on the propagation of TE--polarized light \cite{LL8} in a two--dimensional fish eye and proved perfect resolution for this case. Here the electromagnetic wave equation, the Helmholtz equation (\ref{helmholtz}), is the scalar wave equation in a two--dimensional geometry with $n^2 (\mathrm{d}x^2 + \mathrm{d}y^2)$ as the square of the line element, because the Helmholtz equation can be written as
\begin{equation}
0 = \frac{1}{n^2} \nabla^2 E + k^2 E = \frac{1}{\sqrt{g}}\,\partial_A \sqrt{g}\, g^{AB} \partial_B E + k^2 E
\end{equation}
where the indices refer to the coordinates $x$ and $y$ in a geometry \cite{LeoPhilbin} with metric tensor $g_{AB}=n^2\mathds{1}$, its determinant $g=n^4$ and the inverse metric tensor $g^{AB}=n^{-2}\mathds{1}$; and where we sum over repeated indices. Consequently, the geometry of light established by Maxwell's fish--eye is not restricted to rays, but extends to waves, which may explain why waves are as perfectly imaged as rays. In contrast, perfect imaging does not occur for the TM polarization \cite{LL8} where the magnetic--field vector $\mathbf{H}$ points orthogonal to the plane. In this case, the corresponding wave equation \cite{LL8} for the magnetic field,
\begin{equation}
\nabla \cdot \frac{1}{n^2} \nabla H + k^2 H = 0 \,,
\label{magnetic}
\end{equation}
cannot be understood as the wave equation in a two--dimensional geometry. For a source placed at the origin we find for the fish-eye profile (\ref{maxwell}) the asymptotic solutions $H \sim r^{-4}$ and $H\sim \mathrm{const}$ at infinity, neither of them forming the required logarithmic divergence of a perfect image in two dimensions. This proves that perfect imaging in the two--dimensional fish eye is impossible for the TM polarization where the geometry is imperfect for waves. On the other hand, the three--dimensional impedance--matched Maxwell fish eye perfectly implements the surface of a four--dimensional hypersphere \cite{LeoPhilbin}, a three-dimensional curved space. We expect perfect imaging in this case.

Perfect imaging is often discussed as the amplification of evanescent waves \cite{Pendry}, but this picture does not quite fit the imaging in Maxwell's fish eye that seems solely caused by the geometry of the sphere. Note that there is an alternative, purely geometrical picture for understanding negatively--refractive perfect lenses as well \cite{GREE}: they implement coordinate transformations with multiple images. What seems to matter most in perfect imaging is the geometry of light \cite{LeoPhilbin,SchleichScully}.

\section*{Acknowledgments}
I thank
Lucas Gabrielli, 
Michal Lipson,
Thomas Philbin,
and
Tom\'a\v{s} Tyc
for inspiring discussions.
My work is supported by a Royal Society Wolfson Research Merit Award and a Theo Murphy Blue Skies Award of the Royal Society.


\end{document}